\documentclass[prd,nofootinbib,twocolumn,showpacs]{revtex4}
\usepackage{graphics}
\usepackage{bm}

\begin{document}

\title {Domain Formation: Decided Before, or After the Transition?}

\author{Nuno\ D.\ Antunes}  \affiliation{
Center for Theoretical Physics, University of Sussex, \\
Falmer, Brighton BN1 9WJ, U.K.}
\author{Pedro\ Gandra} \author{Ray\ J.\ Rivers}
\affiliation{ Blackett Laboratory, Imperial College\\
London SW7 2BZ, U.K.}

\begin{abstract}
There is increasing evidence that causality provides useful bounds
in determining the domain structure after a continuous transition.
In devising their scaling laws for domain size after such a
transition, Zurek and Kibble presented arguments in which
causality is important both {\it before} and {\it after} the time
at which the transition begins to be implemented. Using numerical
simulations of kinks in 1+1 dimensions, we explain how the domain
structure is determined exclusively by what happens {\it after}
the transition, even though the correlation length freezes in {\it
before} the transition.
\end{abstract}


\

\maketitle

\section{Introduction}

It is inevitable that causality will provide constraints on the
formation of domains after a continuous transition. Although,
adiabatically, correlation lengths diverge, the finite speed at
which the order parameter can adjust to the changing environment
guarantees that correlation lengths remain finite for transitions
implemented in finite times. Assuming that systems do change as
fast as they can, Kibble \cite{kibble1} and Zurek \cite
{zurek1,zurek2} predicted simple scaling behaviour for the maximum
correlation length of the order parameters of the early universe
and condensed matter systems respectively, as a function of the
cooling rate.

Domain formation, the frustration of the order parameter fields,
is often visible through topological defects, typically vortices,
which mediate between different equivalent ground states. Since
defects are, in principle, observable, they provide an excellent
experimental tool for confirming this scaling behaviour, insofar
as the defect separation can be correlated directly to the order
parameter. Several experiments have been performed to measure
defect density that support the KZ scaling behaviour under this
assumption \cite{florence,roberto,technion2}, which we term the
Kibble-Zurek scenario, or are commensurate with its predictions
\cite{technion2,grenoble,helsinki,carmi,digal}.

 In this paper we give a critical reappraisal of the Kibble-Zurek
(KZ) scenario or, more accurately, scenarios. There are several
different ways to derive causal bounds on defect densities.
Although they agree numerically for simple systems, they differ
conceptually in their assumption as to how the length scale that
determines the defect separation arises. The main distinction is
whether this is set when the system freezes in {\it before} the
transition or when the system relaxes {\it after} the transition
has begun.

In this paper we shall show that despite the fact that the
correlation length freezes in {\it before} the transition, this
value does not determine the density of defects upon their
appearance, contrary to the standard formulation of causal bounds.
As we shall see, the defect density is determined entirely by what
happens {\it after} the transition has begun. This latter point
has been argued by one of us (RR) over many years
\cite{al,karra,ray2,ray1} in the context of analytic
approximations and by Moro and Lythe \cite{moro} using numerical
simulations to supplement similar analytic results but, we feel,
can benefit from the explicit demonstration that we give here. It
is also timely in that the first experiments \cite{newjtj} are
being performed that show that the defect density depends on what
happens {\it after} the transition. We shall turn to these in our
concluding section.

\section{The KZ scenarios}

To be specific, consider a system with critical temperature
$T_{c}$, cooled through that
temperature so that, if $T(t)$ is the temperature at time $t$, then $%
T(0)=T_{c}$.
 \[{\dot{T}}(0)= -T_{c}/\tau _{Q}\]
  defines the quench
time $\tau _{Q}$. The adiabatic correlation length $\xi
_{ad}(t)=\xi _{ad}(T(t))$  diverges near $t=0$ but the true
correlation length $\xi (t)$ remains finite. The standard
formulation of the KZ scenario is that the correlation length
${\bar{\xi}}$ that characterizes the
onset of order is the equilibrium correlation length ${\bar{\xi}}=\xi _{ad}({%
\bar{t}})$ at some appropriate time ${\bar{t}}<0$ {\it before} the
transition is implemented.

The argument goes as follows; at time $t$ there is a maximum speed
$c(T(t)) = c(t)$ at which the system can order itself. In the
early universe this is the constant speed of light, in superfluid
$^4He$, say, the speed of second sound, vanishing at $T=T_c$.
[Whether there is critical slowing down or not is irrelevant.]
>From this we can define a relaxation time $\tau (t) =
\xi_{ad}(t)/c(t)$. The earliest suggestion, due to Zurek, is that,
initially, for $t<0$ some way before the transition, the system
behaves adiabatically. However, the system freezes in (the
'impulse' region) at time $-{\bar t}$, where
 \begin{equation}
 {\bar t}\approx \tau ({\bar t}),
 \label{tbar1}
 \end{equation}
i.e. when the time to the transition matches the relaxation time.
It is proposed that the correlation length ${\bar\xi}
=\xi_{ad}(-{\bar t})$ frozen out at this time will determine the
domain size at the transition.

The argument due to Kibble is even simpler; $\xi (t)$ cannot grow
faster than $c(t)$.  If $\xi (t)\approx \xi_{ad}(t)$ until the
system cannot keep up with the cooling, the time $-\bar t$ at
which the system freezes in is defined by the condition
 \begin{equation}
 {\dot\xi}_{ad}(-{\bar t})\approx c(-{\bar t}).
 \label{tbar2}
 \end{equation}
For simple systems both estimates of ${\bar t}$
  agree, to give the allometric form \cite{zurek2}
  \begin{equation}
 {\bar\xi}_<= \xi_{ad}(-{\bar t})= a_<(\tau_Q)^{\sigma_<}.\label{xibar<}
  \end{equation}
 The scaling exponent $\sigma_<$ depends on the (adiabatic) critical behaviour of the
 system as $T\rightarrow T_c-$, and $a_<$ depends on the microscopic properties of the system.

In addition to considering relaxation times before the event Zurek
also invoked \cite{zurek2} causal horizons {\it after} the event
to obtain causal bounds. At time $t$ the causal horizon in which
the order parameter can be correlated is
\begin{equation}
\xi_c(t)\approx 2\int^t_0 ds\,c(s). \label{xic}
\end{equation}
>From this viewpoint our horizon bound for the earliest time ${\bar
t}$ that we can see defects is when $\xi_c(t)$ becomes larger than
the coherence length $\xi_{ad}(t)$. That is, the causal horizon is
big enough to hold a classical defect,
\begin{equation}
\xi_c({\bar t}) = \xi_{ad}({\bar t}). \label{tbar4}
\end{equation}
The correlation length is then ${\bar\xi}_> = \xi_{ad}({\bar t})$.

 A parallel case that the relevant
time for setting ${\bar\xi}$ is {\it after} the transition can be
made for Kibble's analysis, by saying that the impulse regime will
give way to the adiabatic regime once
\begin{equation}
 {\dot\xi}_{ad}({\bar t})\approx -c({\bar t}),
 \label{tbar3}
 \end{equation}
 and it is this that determines the relevant ${\bar t}$.
 Insofar as $\xi_{ad}(t)$ is symmetric, this gives the same ${\bar
 t}$ as in (\ref{tbar2}).
Again, both estimates (\ref{tbar3}) and (\ref{tbar4}) for ${\bar
t}$
  agree, to give
  \begin{equation}
 {\bar\xi}_> = a_>(\tau_Q)^{\sigma_>},\label{xibar>}
  \end{equation}
  where the scaling exponent $\sigma_>$ depends now on the
(adiabatic) critical behaviour of the
 system as $T\rightarrow T_c-$.

For continuous ${\dot T(t)}$ the scaling exponents $\sigma_< =
\sigma_>$ are equal, when both exist. As a result (\ref{xibar<})
and (\ref{xibar>}) are conflated as ${\bar\xi} =
a(\tau_Q)^{\sigma}$, the KZ {\it scaling law}.

In either formulation Kibble and Zurek then make the second
assumption, that the relevant ${\bar\xi}$ sets the initial scale
of the defect network, with defect separation $\xi_{\rm def}
\approx {\bar\xi}$.

In this paper we shall show that despite the fact that the correlation
length freezes in at ${\bar\xi}$ satisfying
(\ref{xibar<}), this does not fix the density of defects.
Instead, the density is determined by the growth of instabilities after the
transition, which essentially matches (\ref{xibar>}) without
invoking causality directly, although all evolution is,
inevitably, causal.

\section{The Model: Length Scales}

We will consider a simple model in 1+1 dimensions that
has been used as a first step in several studies of defect
formation \cite{lag,us}.  This model is easily amenable to numerical
simulation and leads to the correct scaling laws for the defect
density. In a future publication we will deal with systems with
larger number of spatial dimensions. Nevertheless we 
 expect the results discussed below and the analytical 
estimates behind these to generalize to higher dimensions.

 Specifically,  we consider the following Langevin
equation, describing a time-dependent Ginzburg-Landau (TDGL) field
theory:
\begin{eqnarray}
&&\partial_t^2\phi-\nabla^2 \phi + \alpha^2 \,\partial_t\phi +
m^2(t) \phi - 2 \lambda \phi^3 = \alpha \zeta, \label{eom}
\end{eqnarray}
where $\zeta$ is a Gaussian noise term obeying
\begin{equation}
 \langle\zeta(x',t') \zeta(x,t)\rangle=2 T
\delta(x'-x) \delta(t'-t),\,\,\,\, \langle \zeta(x,t) \rangle=0 .
\end{equation}
 In one dimension $\nabla^2\phi
=\partial^2_x \phi$, $\alpha$ measures the amplitude of the noise,
and its square quantifies the dissipative effects. This relation
between the amplitudes of the noise and the dissipative term
ensure that the fluctuation-dissipation condition is satisfied and
that for very long times the system should eventually reach
thermal equilibrium at temperature $T$.

This TDGL model is motivated empirically for overdamped (large
$\alpha$) condensed matter systems but, even in their idealised
form, we would expect relativistic field theories to have
non-linear noise arising from their environments.  However, we
have shown elsewhere that scaling behaviour is, essentially,
independent of the linearity or non-linearity of the noise and we
stick with (\ref{eom}).


 We start by presenting a simplified analytic argument supporting
the assumption that the correlation length does freeze in before the
 transition, as anticipated by Kibble and Zurek. This will also help
 clarifying  the reason why this scale does not necessarily set the
 value of the final defect density.
 In the simplest version
of the model, the mass term is decreased linearly in time from an
initial positive value $\mu^2$ as
\begin{equation}
m^2(t)=-\mu^2 \frac{t}{\tau_Q},\,\,\,\,\,\,-\tau_Q<t\leq 0
\end{equation}
For  $t <-\tau_Q$ the mass term is set to constant $\mu^2$.
For values of time larger (smaller) than $\tau_Q$ ($-\tau_Q$) the
mass term is set to constant $-\mu^2$ ($\mu^2$).

To understand how the system does freeze in at time $-{\bar t}$,
for $t<0$ non-linear excitations have not had time to grow and it
is sufficient to set $\lambda = 0$ in (\ref{eom}). For reasons of
familiarity it is convenient to extend the (now linear) equation
to three space dimensions for $t<0$. The field correlation
function
 $G(r,t)=\langle\phi ({\vec x})\phi ({\vec 0})\rangle_t$ is
expressible as
\begin{equation}
G(r,t) = \int \frac{d^3k}{(2\pi)^3}\,P(k,t)e^{i{\vec k}.{\vec x}},
\label{G}
\end{equation}
 where $P(k,t)$ is the {\it power} in the fluctuations of
wave-vector ${\vec k}$. The solutions to (\ref{eom}) are given in
terms of Airy functions and their generalisations, and are messy
without being very informative (see \cite{moro}).
It is sufficient, for illustrative  purposes, to restrict
ourselves to the case of strong dissipation, whereby Airy
functions become exponentials.  As we have shown before
\cite{kav}, for strong damping,
\begin{equation}
P(k, t) = \int_{0}^{\infty} d\tau \,{\bar T}(t-\tau/2)\,e^{-\tau
k^{2}}\,e^{-\int_{0}^{\tau} d\tau'\,\,m^2 (t- \tau'/2)},
\label{lgpower}
\end{equation}
up to an irrelevant renormalisation. In turn, this gives $G_0 (r,
t) = $
\begin{equation}
                     = \int_{0}^{\infty} d\tau\,{\bar T}(t-\tau/2)
\,\bigg(\frac{1}{4\pi\tau}\bigg)^{3/2}
e^{-r^{2}/4\tau}\,e^{-\int_{0}^{\tau} d\tau'\,\,\epsilon (t-
\tau'/2)}. \label{lgcorr7}
\end{equation}

It is not difficult to see that, for $r$ large enough,
\begin{equation}
G(r,0)\approx \frac{1}{4\pi r}\exp [a(r/\bar\xi_<)^{4/3}]
\label{longr}
\end{equation}
in this case. In (\ref{longr}) ${\bar\xi}_<$ is, indeed, the
correlation length that follows from (\ref{tbar1}) and
(\ref{tbar2}) above and $a\approx 1$.

 What is important here is
that the correlation length ${\bar\xi}_<$ is determined from the
position of the nearest singularity in the complex $k$-plane of
the integral (\ref{G}) after performing the angular integrations
or, equivalently, from the {\it large-distance} behaviour of the
correlation function.

The reason why, both for this and the more general case of
arbitrary $\alpha$, this length does not set the scale for the
density of defects is that defects are identified by the field
zeroes at their cores when they are formed at $t>0$. Initially,
prior to $t=0$, there are zeroes on all scales. However, once the
transition begins to get under way, the instabilities in the long
wavelength modes of the field lead to exponential growth in long
wavelength amplitudes that orders the fields. Insofar that we can
persist with the linearized system after the transition, the
separation of zeroes $\xi_{zero}$ at the formation of defects is
\cite{halperin,maz}
\begin{equation}
\frac{1}{\xi_{zero}^2}=\frac{-1}{2\pi}\frac{G''(r=0,{\bar
t})}{G(r=0,{\bar t})}, \label{shortr}
\end{equation}
where dashes denote differentiation with respect to $r$. That is,
the information comes entirely from the {\it short distance}
behaviour of $G(r,t)$ for $t>0$, rather than the long-distance
behaviour that determined ${\bar\xi_<}$ of (\ref{longr}) for
$t\leq 0$. It is determined by the second moment of the power
spectrum, rather than its analytic structure.

It happens that, in mean-field theory at least, $\xi_{zero}$ of
(\ref{shortr}) can have the same scaling behaviour as
${\bar\xi}_>$ of (\ref{xibar>}), that would be obtained from
(\ref{tbar3}) and (\ref{tbar4}). There are only logarithmic
corrections to the scaling laws, provided thermal fluctuations,
that introduce a further length scale (the Ginzburg correlation
length) are unimportant \cite{ray2}. The work of \cite{moro} shows
just how reliable this analytic approximation is, provided
transients are taken into account.

\section{Numerical Simulations}

For this type of quench, we expect that some time after $\tau_Q$
the field should settle to a series of alternating positive and
negative
 vacuum regions.
These regions will be separated by the topological defects of the
 theory, kinks and anti-kinks interpolating between the
opposite vacua. The defects will then enter a regime of
slow evolution with pair annihilation taking place at long times.
During this stage of the evolution there is a one-to-one correspondence
between kinks and anti-kinks and zeros of
the scalar field $\phi$, making it trivial to identify them numerically.

Depending on the value of the dissipation coefficient $\alpha^2$,
we expect that the final density of defects $N_{\rm def}=
1/{\xi_{\rm def}}$ will scale with  different powers of the quench
time scale. For high dissipation the system is effectively
first-order in time, whereas for low $\alpha$ the second
derivative dominates the evolution and the system behaves in a
relativistic fashion. Assuming that the final defect density is of
the order of the inverse of the freeze-out correlation length,
$N_{\rm def}$ should scale as $\tau_Q^{-\sigma}$. As discussed
extensively in the literature and verified numerically for this
model \cite{lag,us}, the exponent is given by $\sigma\simeq1/4$ in
the dissipative regime, and $\sigma\simeq 1/3$ in the relativistic
case.

We will now look at variations of the quench model described above
in which we modify the basic system so that the properties of the
quench are qualitatively different before and after the transition
takes place at $t=0$. By measuring the final defect densities and
determining the corresponding scaling powers, we will be able to
determine which period of the evolution dictates the outcome.
Though transitions such as envisaged here are unlikely to be
realisable in a physical situation, such scenarios are
useful as idealized experiments. We stress that our primary concern here
is with a matter of principle, given the general acceptance of the KZ
scenario in its original form. Nevertheless, as we will discuss below,
experimental results in transitions in Josephson Junctions can
be understood in terms of distinct behaviours of the system before
and after the transition.

As already mentioned, the value of the dissipation in equation
(\ref{eom}) determines the scaling power of the final density of
defects. In the first variation we will rely on this fact to
discriminate between the effects of the two main stages of the
quench on the final outcome. The idea is to allow $\alpha$ to take
two distinct values, $\alpha = \alpha_<$ for $t<0$ and $\alpha
=\alpha_>$ for $t>0$, characterizing the dissipation and the
thermal noise before and after the transition respectively. If
$\alpha_<$ and $\alpha_>$ are values typical of different regimes
(under and over-damped and vice-versa) for $t<0$ and $t>0$
respectively, the measured value of $\sigma$ should reflect the
relevant period of the evolution.

 The numerical procedure used in this work follows very closely that of
\cite{us}. We will outline it here briefly and refer the reader to
the aforementioned publication for a more detailed description.
The parameters of the potential are $\mu^2=1.0$ and $\lambda=1.0$
and the bath temperature $T$ is set to a low value, typically
$T=0.01$. For every fixed choice of $\alpha_<$ and $\alpha_>$  a
set of quenches is performed with $\tau_Q=2^n$,  $n=1,2,..,9$ and the
final defect density is obtained by counting zeros of the field at
a final time, defined as a multiple of $\tau_Q$. This approach
to determining the final kink number leads to slight discrepancies
in the estimate of $\sigma$ in the extreme high/low-dissipation cases.
These are nevertheless easily controlled \cite{lag,us} and as we will
see below the accuracy obtained is more than sufficient for the
needs of the present work.
 Finally, we perform a fit of the final
defect density versus $\tau_Q$ to a power law and thus obtain the
scaling exponent $\sigma$.

\begin{figure}
\scalebox{0.45}{\includegraphics{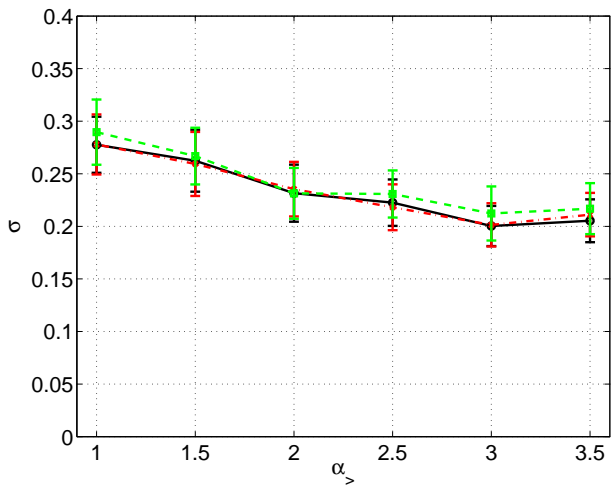}} \caption{Final defect
density scaling power as a function of the noise amplitude for
$t>0$. The amplitude before the transition was set to
$\alpha_<=0.4$ and $\alpha_<=0.5$ for the full and dash-dotted
curves respectively. For comparison we plot (dashed curve) the
standard result for a case with constant dissipation throughout
the quench, {\it i.e.} $\alpha_<=\alpha_>$. The error bars are the
standard deviation of the result over a set of 10 independent
realizations of the quench.} \label{fig1}
\end{figure}

 In Fig.~\ref{fig1} we show the results for a case where the evolution
 before the transition is well in the under-damped regime. In \cite{us}
 we showed that for this specific model, the transition between the
 two types of evolution takes place at about $\alpha_c=0.8$. The two values
 of $\alpha_<=0.4,\, 0.5$ used here are both comfortably below
 $\alpha_c$.  In both cases we allow $\alpha_>$ to vary between $1.0$
 and $3.5$, covering a part of the over-damped region of parameter
 space. As illustrated in Fig.~\ref{fig1} the results show clearly that
 the scaling parameter $\sigma$ is determined exclusively by the value of
 $\alpha_>$. The result for the scaling exponent is unmistakably
 typical of the over-damped regime, despite the fact the part of the evolution
 was under-damped. Not only that, but the precise value of $\sigma$
 does not seem to affected by $\alpha_<$.
 Within the error bars, the curves obtained are indistinguishable
 from the case where the dissipation is kept constant, and equal to $\alpha_>$,
 throughout the transition.

\begin{figure}
\scalebox{0.45}{\includegraphics{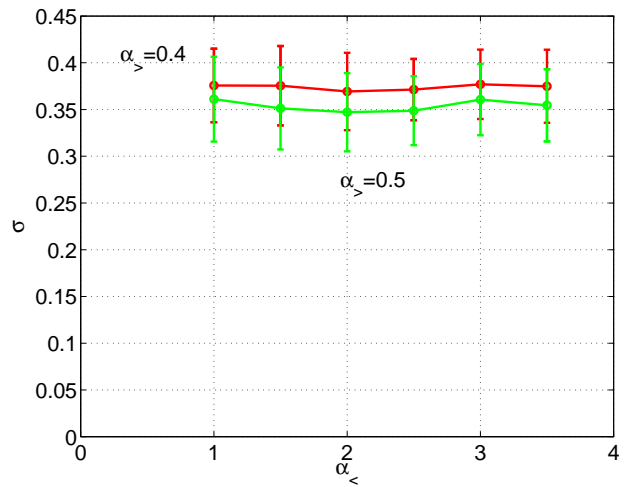}} \caption{Defect
scaling power as a function of the
 noise amplitude for $t<0$. For $t>0$ the amplitude are
 $\alpha_>=0.4$ and $\alpha_>=0.5$. Error bars are as in Fig.~\ref{fig1}.}
\label{fig2}
\end{figure}

 In Fig.~\ref{fig2} we show the results for the opposite situation.
 Here we kept the dissipation {\it after} the transition constant
 and in the under-damped regime, while keeping $\alpha_<$ in the
 over-damped region. The measured $\sigma$ is constant within
 error bars, and has a value typical of an under-damped evolution.
 The values of the scaling power for a quench with constant $\alpha$
 set to $0.4$ and $0.5$ are $\sigma=0.39$ and $\sigma=0.35$ respectively
\footnote{
 Since we are working on the region of very low dissipations
 these are slightly larger than $1/3$, the typical value for
 the under-damped regime, a consequence of saturation phenomena
  \cite{lag,us}.}.
 As in the previous example, the values shown in Fig.~\ref{fig2}
 coincide with these.
 Once again this strongly suggests that the outcome is determined
 exclusively by the period of the quench taking place after the
 phase transition.

For our second example we take a more radical departure from the
original model (\ref{eom}). While maintaining the dissipation at a
 constant value throughout the whole evolution, we change the rate
of the quench at $t=0$. For negative times the squared mass term
$m^2(t)$ will go from $\mu^2$ at $t=-\tau_{Q<}$ to zero at $t=0$.
For $t>0$ an alternative quench rate is introduced, with $m^2(t)$
decreasing linearly and reaching $-\mu^2$ for $t=\tau_{Q>}$. As in
the previous section, we expect the outcome to reflect the
relative importance of the two periods of the evolution.

\begin{figure}
\scalebox{0.45}{\includegraphics{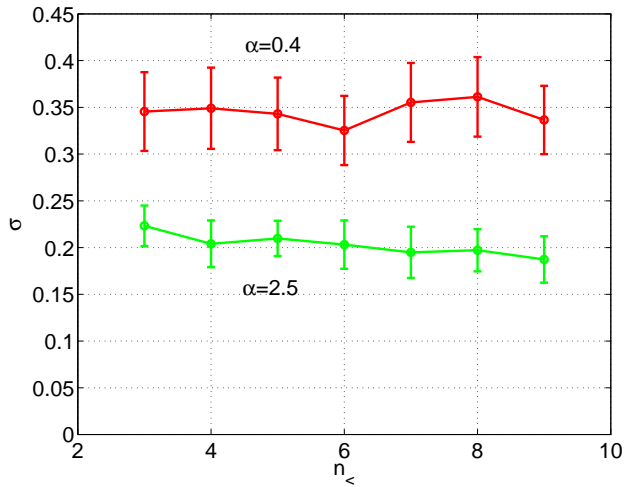}} \caption{Defect
scaling power as a function of the fixed, $t<0$ quench time
 $\tau_{Q<}=2^{n_<}$, for under-damped ($\alpha=0.4$) and over-damped
 ($\alpha=2.5$) cases. Error bars are as in Fig.~\ref{fig1}.}
\label{fig3}
\end{figure}

In Fig.~\ref{fig3} we show the results for two cases, with
$\alpha$ in the under-damped and over-damped region respectively.
For both sets of simulations, $\tau_{Q<}$  was kept constant at a
value $\tau_{Q<}=2^{n_<}$ for fixed $n_<$ whilst
 $\tau_{Q>}=2^{n_>}$ was
varied throughout the usual range, with $n_>=1,\dots 9$. If the
final number of defects were determined by $\tau_{Q<}$ we should
observe no scaling, with a constant number of defects being
obtained at the end of each simulation. Instead, we found that the
final density clearly scales with $\tau_{Q>}$, with a power that
is determined by the magnitude of $\alpha$. As seen in
Fig.~\ref{fig3} the value of the scaling power changes very little
with the quench rate at negative time. For the two choices of
$\alpha=0.4, 2.5$, the exponents for the basic quenches with
$\tau_{Q<}=\tau_{Q>}$ are $0.39$ and $0.23$ respectively. The
measured $\sigma$'s lie very near these, confirming that the $t>0$
part of the evolution is the dominant one. As an extra check we
simulated the reversed situation where $n_>$ was kept fixed and
$n_<$ was allowed to vary in the usual range. As expected the
final defect density was constant within error bars, showing no
dependence on the value of $\tau_{Q<}$ for negative times. Once
again this highlights the fact that the mechanism of defect
formation is dictated by what happens after the transition takes
place, in variance with some of the arguments of
\cite{kibble1,zurek1,zurek2}.

\section{Conclusions}

If there had been any doubt before, we have shown conclusively
that defect density is controlled by the cooling of the system
after the critical temperature has been passed.  It is true that
correlation lengths do freeze in before the transition. However,
their defining long-distance behaviour is irrelevant to defect
densities, which are controlled by the short-distance behaviour
that determines field zeroes.

As we have stressed, transitions with discretely different
behaviour before and after the critical temperature of the type
that we have examined are unlikely to have direct physical
counterparts. However, the transitions in Josephson Junctions
provide a more subtle example where behaviour before and after the
transition is very different. Before the conductor-superconductor
transition we have two bulk conductors, individually amenable to
the Zurek analysis of freezing in of correlation lengths
\cite{zurek2}. For these, (\ref{xibar<}) is appropriate, with
$\sigma_< = 1/4$. However, after the transition we have the
Josephson effect, for which field ordering is not determined by
the behaviour of the individual superconductors, for which
(\ref{xibar<})  would have given an identical $\sigma_> = 1/4$
\cite{zurek2}. Instead, field ordering is constrained  by the
velocity of light in the oxide, the Swihart velocity
\cite{KMR&MRK}. Naively, we could not easily distinguish between
the two possibilities since, for idealised symmetric junctions
this again gives the same scaling behaviour $\sigma_> = 1/4$
\cite{KMR&MRK} after the transition, albeit with a different
prefactor. However, for realistic junctions proximity effects and
only partial critical slowing-down $\sigma_< = 1/4$ enforce
$\sigma_> = 1/2$ \cite{newjtj}. Empirically, the data is described
well by the latter \cite{newjtj}, and is not compatible with
$\sigma_< = 1/4$.

Both the analytical estimates and the numerical simulations
presented above suggest strongly that our results should apply to
domain formation in generic second-order phase-transitions.

 N. D. Antunes was funded by PPARC. P. Gandra was supported by FCT,
 grant number PRAXIS XXI BD/18432/98. R. J. Rivers would like to acknowledge support
 from the ESF COSLAB programme.

\end{document}